# Barrier Coverage in Mobile Camera Sensor Networks with Grid-Based Deployment

XIAO-LAN LIU[1], BIN YANG[2] AND GUI-LIN CHEN[3]
*School of Computer and Information Engineering,*
*Chuzhou University*
*Chuzhou, 239000 P.R. China*
*E-mail: {[1]lxl; [3]glchen}@chzu.edu.cn*
*{[2]yangbinchi }@gmail.com*

Barrier coverage is a critical issue in wireless sensor networks for many practical applications, e.g., national border monitoring, security surveillance and intruder detection, etc. Its aim is to detect intruders that attempt to cross the protected region. Available works mainly focused on the barrier coverage of directional sensors (e.g., cameras), which have limited coverage area and sensing angles. In this paper, we study how to efficiently improve barrier coverage using mobile camera sensors, where camera sensors are deployed by a grid-based strategy. We first propose a novel full-view covered model of mobile camera sensors. With this model, we divide the target field into connected grids and deploy mobile camera sensors for each grid. Then we construct a weighted directed graph to model the full-view covered grids and their relationship. Based on the graph, we employ Dijkstra's algorithm to obtain a shortest camera barrier, which is a connected full-view covered zone across the target field and contains minimum number of camera sensors. Finally, extensive simulation results are provided to illustrate that our proposed solution outperforms others in terms of coverage probability and minimum number of camera sensors.

## 1. INTRODUCTION

Barrier coverage is a fundamental issue for supporting various sensor network applications such as battlefield surveillance, critical resource protection, border monitoring, and airport intruder detection. It depicts the capability to detect intruders who attempt to pass through the protected areas. A barrier in wireless sensor network is composed of a set of sensors such that any intruder can be detected by at least one sensor in the barrier when the intruder crosses the region from one side to the opposite side [1]. In comparison with other coverage types (e.g., point coverage that covers specific points of interest and area coverage that covers the entire region [2]), the barrier coverage can effectively reduce the number of sensors required, so it has been attracting much attention in practice.

The barrier coverage in previous works mainly focused on traditional scalar sensors, which are used for measuring scalar physical phenomena from the surrounding environment. Recently, many works have been dedicated to the study of directional sensors (e.g., cameras) barrier coverage [3]-[7], where camera can retrieve much richer information such as images or videos from the physical world in compared with traditional sensors. Barrier coverage in camera sensor networks is first introduced in [7], where the camera coverage is a fan area and each camera sensor from different positions may generate very different views of the same object. As a result, the camera sensor has the ability to fully cover the object, but it may not view its face image. Later, literature [5] proposed the model of full-view coverage, where an object is considered to be full-view covered if no matter which direction the object faces, there is always a camera sensor so that the object is within the sensor's range and the sensor's viewing direction is sufficiently close to the object's facing direction. Based on the model of full-view coverage, Yi Wang [3] proposed a method to construct a full-view camera barrier, where any intruder can be identified once it crosses the barrier no matter which direction it faces. To further reduce the number of needed cameras for barrier coverage, the Minimum Camera Barrier Coverage Problem is proposed in [1], where the authors improve the Shortest Path selection algorithm proposed in [3].

However, these works as mentioned above mainly consider static camera barrier coverage. After the initial deployment, it is difficult to improve barrier coverage for limited coverage area and sensing angles. Especially for some application scenarios, camera sensors have to be randomly deployed in a hostile environment or a hard-to-reach region. As a result, it can lead to a relatively large number of redundant cameras which are not used properly. Moreover, the cost of camera sensors is fairly high. Fortunately, with recent technical advances, a lot of works have been done on barrier coverage using mobile sensors. However, most of existing works mainly focus on omni-directional sensor [8]-[15]. Until in the literature [16], Wang et al. first presented to study barrier coverage with mobile directional sensors. Since then, they in [17] further studied how to efficiently use mobile directional sensors to achieve k-barrier coverage. It is notable that this is a seminal work regarding k-barrier coverage with mobile directional sensor. However, this work does not consider camera full-view coverage, which has significant impact on the quality of coverage.

In this paper, to take advantage of the camera full-view covered model presented in [5], we study how to efficiently improve barrier coverage with mobile camera sensors, and the sensors are deployed by a grid-based strategy [18], which is used for simplifying barrier coverage area complexity after the initial random deployment. The main contributions of this paper are summarized as follows: First, we propose a novel full-view covered model, where mobile cameras can achieve full-view coverage for the grid-based area. With this model, we divide the original continuous target space into connected discrete grid-based spaces which can satisfy the condition of full-view coverage, and deploy mobile camera sensors for each grid. Then we develop a weighted directed graph to model the full-view covered grids and their relationship. Furthermore, we employ Dijkstra's algorithm to obtain a shortest camera barrier, which consists of some connected full-view covered grids and contains minimum number of camera sensors. Finally, we validate our results through extensive simulations.

The rest of this paper is organized as follows. Section II reviews the related work. Section III defines some notations and related coverage model. Section IV gives the detailed description on mobile camera full-view covered detection. Based on this model, In Section V we divide the target field into connected full-view covered grids. Then, Section VI gives the detailed description on how to select minimum camera sensors to form a camera barrier in the grid-based deployment networks. The simulation results are provided in Section VII. Finally, Section VIII concludes the paper.

## 2. RELATED WORK

Extensive studies have been conducted on coverage problem in wireless sensor networks (WSNs), this problem can be classified into three categories: point coverage, area coverage and barrier coverage [2]. Barrier coverage is to detect intruders which penetrate a protected region, and barrier coverage in WSNs is first introduced in the context of robotic sensors [19], where it concerns a sensor network's capability to detect intruders crossing the barrier. Since the seminal work of [19], barrier coverage problems in WSNs based on isotopic sensing model have been extensively studied [20]-[24]. Kumar et al. [21] firstly defined the notion of k-barrier coverage for WSNs and introduced two notions of probabilistic barrier coverage - weak barrier coverage and strong barrier coverage. In [22], Liu and Towsley studied the barrier coverage problem on two-dimensional plane and two-dimensional strip sensor networks using percolation theory. The literature [23, 24] first presented barrier coverage of three dimensional sensor networks. Some other studies are to explore the effects of different sensor deployment strategies and mechanisms to improve barrier coverage [18, 25, and 26]. Most of the early studies assume that the sensor locations follow a Poisson point process where sensors are distributed in a large area uniformly at random. Saipulla et al. [25] first studied the barrier coverage of the line-based deployment rather than the Poisson distribution model. He et al. [26] designed curve-based sensor deployment algorithms for barrier coverage. Wang et al. [18] proposed the grid-based deployment strategy for sensor coverage. The aforementioned works focus on traditional scalar sensor which sensing range is an isotropic circular.

Recently, barrier coverage in directional sensor networks has gradually received more and more attention. Adriaens et al. in [27] first presented an optimal polynomial time algorithm for computing the worst-case breach coverage in directional sensor networks. Barrier coverage in camera sensor networks was introduced in [7]. To deal with the special requirement of directional camera sensors, the definition of full-view coverage is introduced in [5]. Based on the model of full-view coverage defined in [5], the literature [3] proposed a method to construct a full-view camera barrier from an arbitrary deployment. To further reduce the number of needed cameras for barrier coverage, the authors [1] improve algorithm of literature [3] and propose minimum camera sensors path selection algorithm in wireless camera sensor networks.

With the development of mobile sensors, sensor mobility is exploited to improve barrier coverage [8]-[15]. Shen et al. [9] studied the energy efficient relocation problem for barrier coverage in mobile sensor networks. Ban et al. [10] studied the problem on how to relocate mobile sensors to construct $k$ grid barriers with minimum energy consumption. Keung et al. [11] focused on providing $k$-barrier coverage against moving intruders in mobile sensor networks. Saipulla et al. [12] proposed a greedy algorithm to find barrier gaps and moved mobile sensors with limited mobility to improve barrier coverage. Since then, they in [13] further studied the barrier coverage of a line-based deployment strategy and exploit sensor mobility to improve barrier coverage. He et al. [14, 15] studied barrier coverage in sensor scarcity case by dynamic sensor patrolling. The above works mainly focus on omni-directional sensors, while neglecting to use mobile directional sensors to form barrier coverage.

Recently, some initial works have studied how to use directional sensor rotation and mobility to construct barrier coverage. Tao et al. [28, 29] investigated strong barrier coverage using directional sensors, where sensors have arbitrarily tunable orientations to provide good coverage. Wang et al. [16] first studied barrier coverage in hybrid directional sensor networks with both stationary and mobile sensors. They in [17] further studied how to efficiently use mobile directional sensors to achieve k-barrier coverage. However, this work neglects camera full-view coverage, which has significant impact on the quality of coverage.

Compared to these aforementioned works on barrier coverage, our study considers comprehensively three factors: mobile camera sensors, full-view barrier coverage, and grid-based deployment model. By taking of full-view covered mode and grid-based deployment strategy, we study how to efficiently improve barrier coverage and reduce camera sensors number using mobile camera sensors.

# 3. MODEL AND NOTATIONS

Camera sensors are deployed randomly to monitor a bounded target region $R$. The cameras can move and rotate within a predefined range. We use $S$ to denote the set of $n$ camera sensors deployed in $R$. Each camera $S_i$ has a sensing range $r$, a field-of-view angle $\varphi$ and an orientation vector $\overrightarrow{f_i}$, which together define the sensing sector as shown in Figure.1 (a). Each camera is aware of its own location and its communication distance is twice the sensing radius of $r$. So if camera sensors $S_i$ and $S_j$ can communicate with each other, the distance $||S_iS_j||$ between them must satisfy relationship: $||S_iS_j||<=2r$.

**Definition 3.1** (Coverage) As shown in Figure 1(b), a point $P$ is covered by a camera sensor $S_i$ if $P$ is within the sensing sector of $S_i$, $\left\|PS_i\right\|<r$ and $a(\overrightarrow{f_i},\overrightarrow{S_iP})<\varphi/2$, where $a(\overrightarrow{f_i},\overrightarrow{S_iP})$ denotes the angle between them, $\overrightarrow{S_iP}$ is the vector from $S_i$ to $P$.

**Definition 3.2** (Full-View Coverage) As shown in Figure.1(c), a point $P$ is full-view covered if for any facing direction $\overrightarrow{d}$, there is a camera sensor $S_i$, such that $P$ is covered by $S_i$ ,and the camera's viewing direction $\overrightarrow{PS_i}$ is sufficiently close to the point's facing direction, and the relationship should hold: $\alpha(\overrightarrow{d},\overrightarrow{PS_i})\le\theta$. Here $\theta\in$ [0, $\pi/2]$ is a predefined parameter which is called the effective angle. A region is full-view covered if every point in it is full-view covered.

Based on the model, we introduce the following definitions for barrier coverage.

**Definition 3.3** (Barrier Coverage) For a given bounded area $R$, If there exists any line which passes through the area from the side to the contrary side and the line is the full-view coverage, the region is called barrier coverage area.

**Definition 3.4** (Camera barrier Coverage). For a given bounded area $R$, with one side being the entrance and the opposite side being the destination. A camera barrier $C$ consists of connected camera barrier coverage areas and all the crossing paths from one point on the entrance side to the destination side intersects with $C$.

Based on the definitions of camera barrier coverage, we find a camera barrier $C$ which contains minimum number camera sensors for the given area $R$ in the following contents.

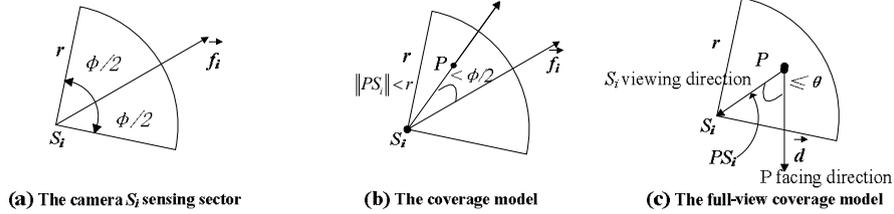

**(a)** The camera $S_i$ sensing sector        **(b)** The coverage model        **(c)** The full-view coverage model

**Figure 1: The camera sensor $S_i$ coverage model.**

# 4. MOBILE CAMERA FULL-VIEW COVERED MODEL

Based on full-view covered model proposed in [5], we present mobile camera full-view covered model in this section. The purpose of the proposed model is to efficiently improve the quality of barrier coverage using mobile camera sensors and reduce the number of needed mobile camera sensors in the full-view coverage. We show the details in the subsequent section.

## 4.1 Background and Motivation

To study mobile camera full-view coverage problem, our solution is motivated by the concept of full-view coverage [5], which is defined as follows: if point $p$ is full-view covered there exists a camera $s$ to cover it no matter which direction it faces and the camera's viewing direction is sufficiently close to the point's facing direction. The model in [5] uses static camera sensors to achieve full-view coverage, which wastes a lot of cameras. Simultaneously, the cost of each camera sensor is fairly high. So optimizing the number of cameras used in full-view coverage is desirable.

In following scenarios, we propose a novel model called mobile camera full-view coverage model. We take advantage of mobile sensors [16, 17] to provide the required full-view coverage and improve the deterministic deployment pattern in [3]. One fundamental difference between them is that we add mobile camera to conduct full-view coverage. According to definition of camera full-view coverage, we transform the monitored full-view coverage area into line full-view coverage area, so the real question is how to deploy mobile camera sensors to achieve full-view line coverage. We show the details in the following deployment. Firstly, we describe the deployment pattern and then analyze the number of cameras used in various deployment parameters.

## 4.2 Description of Deployment Pattern

We first place camera sensors one by one along a line above the barrier with distance $h$ to it, where $h$ is a parameter to be defined later. On this line, any two adjacent deployment spots are separated by distance $\delta$ (to be defined later). At each deployment spot, the camera's orientation vector $f_i$ points down to the barrier and swings with angle $\alpha$ along this direction (to be defined later). Note that each camera can be considered to cover a field-of-view angle $\alpha$. Symmetrically, we place another set of camera sensors along a line with distance $h$ below the barrier. Each camera sensor on this line points up to the barrier and swings with angle $a$ along this direction (Figure 2(a)).

Now we elaborate how to derive the above three parameters: $h$, $\alpha$ and $\delta$. Given the camera's parameters($r$, $\varphi$, $\theta$), we have some flexibility in choosing one parameter from ($h, \alpha, \delta$), and the choice of the other two depends on the chosen one. To simplify the model, we set $45° \leq \theta \leq 90°$. We give the relationship among them in the following theorem 4.1.

**Theorem 4.1** (Deployment conditions) Given $0 \leq h \leq r$, $45° \leq \theta \leq 90°$, in order to guarantee that every point of the barrier is full-view covered, three parameters should satisfy the relationship: $tan(\alpha/2) \leq \delta/h$. Evenly, when the two are equal, the number of needed cameras for full-view coverage is the most small.

Proof: In the first case, we suppose the claim is incorrect. When $tan(\alpha/2) > \delta/h$, every point of the barrier is full-view covered. So we choose a point $p$ randomly, as shown in Figure 2(b). The point $P$ is considered as the intersection between the camera's orientation vector and the barrier. According to the definition 3.2, if the point $P$ is full-view covered, the angle $\theta$ satisfies the relationship: $\theta \in [0, \pi/2]$. However, if the orientation of the point $P$ is parallel to the direction the barrier, the minimum angle between a camera sensor $S$ and the point $P$ is $\pi/2$. By the induction method, the theorem holds. In the second case, the "minimum cameras" part is obvious.

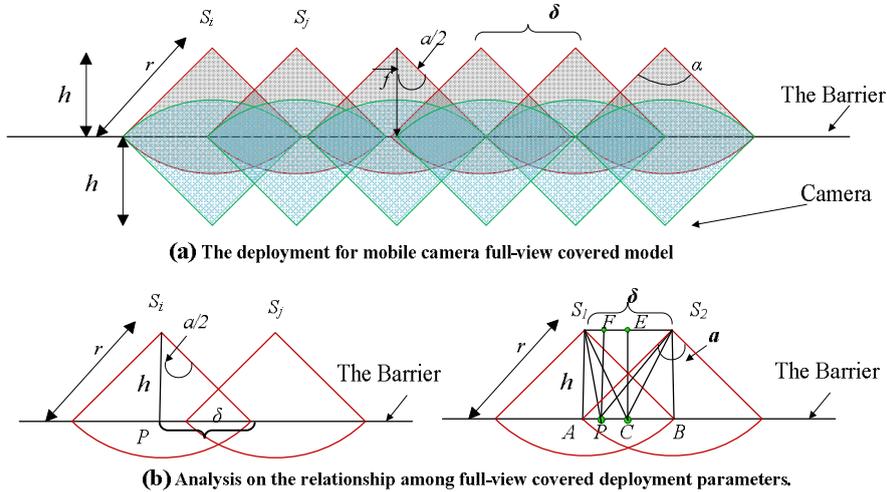

**(a)** The deployment for mobile camera full-view covered model

**(b)** Analysis on the relationship among full-view covered deployment parameters.

**Figure 2: The deployment and relationship description for mobile camera full-view covered model.**

### 4.3 Detection Method Description

Based on the above discussion, we have the following method description regarding full-view covered model detection. Firstly, the barrier can be partitioned into sub-lines by making use of camera sensors' projection in the barrier, where each sub-line is uniform. Then we transform the whole full-view covered line into full-view covered sub-lines. Finally, based on the definition 3.2, we only need to verify if the condition holds for every point in each sub-line.

**Lemma 4.2** (Sub-line Condition) The barrier is full-view covered if and only if the every sub-line is full-view covered by the given set of camera sensors.

Proof: The claim is obvious. The barrier consists of some sub-lines, each of which is similar, and is covered by a set of unified camera sensors. Thus the claim is proved.

If we verify that the sub-line is full-view covered, the trickiest part is to determine if every point on the sub-line is full-view covered. However, there are still infinite numbers of positions to consider. In the subsection, we only need to verify a given sub-line $AB$. If $AB$'s midpoint $C$ is full-view covered, then $AB$ is full-view covered.

**Lemma 4.3** (Midpoint Condition) The sub-line $AB$ is full-view covered if and only if the midpoint $C$ of sub-line $AB$ is full-view covered by the given set of camera sensors.

Proof: As shown in Figure 2 (c). Given the midpoint $C$, a random $P$, sub-segment $AB$, camera $S_1$, $S_2$, the point $C$'s projection on the segment $S_1 S_2$ is $E$, and the point $P$'s projection on the segment $S_1 S_2$ is $F$. In the proof, we only consider points in two directions: one direction is parallel to the barrier, and we describe it as vector $\overline{CB}\ or\ \overline{PB}$. The other is perpendicular to the barrier, and we describe it as vector $\overline{CE}\ or\ \overline{PF}$. As long as the two directions stand, the opposite directions are symmetrical, the conclusion is also obvious. For the other directions, we prove them in the following equation. In the triangle $\triangle BCS_2$, if the angle $\angle S_2CB<\theta$, there is $\angle S_2PB<\theta$ in the triangle $\triangle S_2PB$. In the vertical direction of the barrier, if the angle $\angle ECS_1<\theta$ in the triangle $\triangle ECS_1$, there is $\angle FPS_1<\theta$ in the triangle $\triangle FPS_1$. Even if let the point $P$ closes to the limit $A$, the above conclusion is also obvious.

To this end, we make use of an equivalent condition on full-view coverage proposed in [5]. A given point $p$ is full-view covered if the angle $a$ between any connected two cameras and the point $p$ satisfies the relation: $\alpha\leq 2\theta$. To simplify the model, we set $45°\leq\theta\leq 90°$. All points in the sub-line are located within range of the sensing radius $r$.

**Theorem 4.4** (Model relationships) Given $0\leq h\leq r$, $45°\leq\theta\leq 90°$, $tan(\alpha/2)=\delta/h$, where the formula is given in the theorem 4.1. In order to guarantee that every point of the sub-line is full-view coverage, the minimum value for $\delta$ is $\delta\leq 2h\cdot tan(\theta)$, the minimum value for $\alpha$ is $\alpha\leq 2\cdot arctan(\delta/h)$, and furthermore, $h$ should be smaller than $h_0=r\cdot sin(\theta)$.

Proof: As shown in Figure 2(c). In the proof, we only consider points in two directions: one direction is parallel to the barrier, and we describe it as vector $\overline{CB}$. The other is perpendicular to the barrier, we describe it as vector $\overline{CE}$. As long as the two directions stand, we know that the other directions also stand through the lemma 4.3. Considered to vertical direction of the barrier, if a point $C$ is covered, there is $\angle S_1CS_2\leq 2\theta$. In other words, the angle $\angle ECS_2$ satisfies $\angle ECS_2\leq\theta$. In this condition, there exists $\delta/2h\leq tan(\theta)$, so the minimum value for $\delta$ is $\delta\leq 2h\cdot tan(\theta)$. As for the minimum value $\alpha$, it has been proved in the theorem 4.1. In connection with $h$, in order to ensure that the point $A$ is covered in the triangle $\triangle ABS_2$, the angle $\angle S_2AB$ must satisfy $\angle S_2AB\leq\theta$, in other words, there is $h/r\leq sin(\theta)$, therefore, $h$ should be no bigger than $h_0=r\cdot sin(\theta)$.

Similarly, considered to the barrier direction, if a point $C$ is covered, there is $\angle S_2CB\leq\theta$. In this condition, there exists $2h/\delta\leq tan(\theta)$. We obtained $\delta\leq 2h\cdot tan(\theta)$ from the theorem 4.4. So the relationship $2h=\delta$ is obtained. Through the above analysis, the relationship $\delta=r\cdot 2\sqrt{5}/5$, $h=r\cdot\sqrt{5}/5$ and $\alpha=2\cdot arcsin2\sqrt{5}/5$ are obtained respectively. We give the optimal relationship among them in the following theorem 4.5.

**Theorem 4.5** (Optimal model) Given $0\leq h\leq r$, $45°\leq\theta\leq 90°$, in order to guarantee that every point of the barrier is full-view covered, three parameters should satisfy the relationship $\delta=r\cdot 2\sqrt{5}/5$, $h=r\cdot\sqrt{5}/5$, $\alpha=2\cdot arcsin2\sqrt{5}/5$.

## 4.4 Analysis on the Number of Cameras

Given the above restrictive relations for $\alpha$, $\delta$, $h$ on $r$, we optimize the relationships such that the total number of cameras used in the full-view coverage is minimized. There are two groups of cameras: the first group consists of cameras deployed on the up-line with distance $h$ to the barrier and the second group consists of cameras deployed on the down-line with distance $h$ to the barrier. Without loss of generality, let us consider a unit length of the barrier and assume all the other parameters are unified. The number of cameras used in a unit length of the barrier $1/\delta$, which depends on the choice of $r$ as indicated in theorem 4.5. In order to minimize the total number of cameras, we need to maximize $r$.

**Theorem 4.6** (Camera number) Given $0\leq h\leq r$, $45°\leq\theta\leq 90°$, where $\delta$ is given in theorem 4.5, the density of cameras needed (i.e., number per unit of length) in the above deployment is $1/\delta$, where $\delta=r\cdot 2\sqrt{5}/5$.

By calculating the numerical value of $1/\delta$, we know that $1/\delta$ is an increasing function of $r$. Figure 3 is an illustration of the above result. It first shows how many cameras are needed to construct mobile camera full-view coverage with length of *100m* when *r* is from *2m* to *10m* and *θ=π/4, π/3* and *α=π/3* respectively. Simultaneously, we compare the number of mobile camera needed for full-view coverage and deterministic deployment proposed in [1]. Then we show that our algorithm reduces the number of mobile cameras used in the full-view coverage.

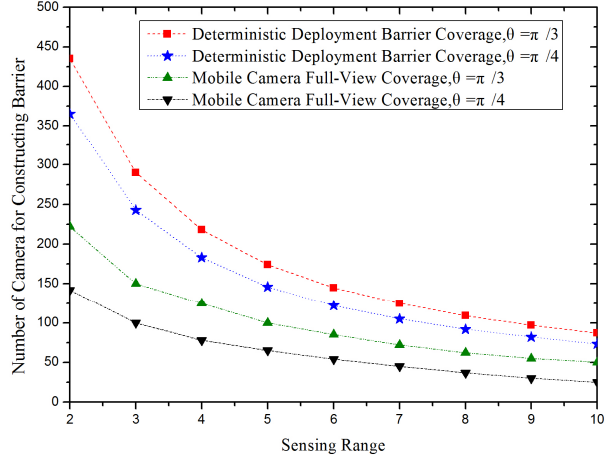

**Figure 3: Number of cameras for constructing barrier vs. Sensing range**

## 5. GRID-BASED DEPLOYMENT STRATEGY

To simplify the complexity for barrier coverage area, we propose the grid-based deployment strategy. In other words, we divide the target field into connected grids. The details are as follows: We first introduce the grid deployment. Then we obtained the relationship of the grid side based on mobile camera full-view coverage model. Finally, we explain how to move camera sensors to achieve the grid-based barrier coverage.

### 5.1 Grid Deployment Initialization

In order to take full advantage of the redundant cameras to achieve full-view coverage, we assume that camera sensors can move and rotate within a predefined range. Each camera has a unique ID and is aware of its own location and the boundary information including the coordinates of four points of *R* [31]. However, it is hard for a single camera to independently decide whether its movement will realize full-view coverage. To make such a decision, the camera sensor requires information about whether it needs to move or not. A grid-based deployment strategy is a natural solution for this problem. So we can divide the target field into grids as shown in Figure 4.

According to the partition, each grid is assigned with a coordinates, and each camera sensor is aware of the coordinates of the located grid. Based on the neighboring information, each camera sensor can derive the number of camera for its own grid. We divide camera sensors in the grid into two types: scale camera and cluster (grid) head. Since many existing techniques on cluster (grid) head [32, 33, and 34] can be directly applied, we will not address these issues in the paper. Other cameras in the grid are collectively known as scale cameras except for the cluster (grid) head.

The rules for assigning coordinates are described below: Each grid is indexed by a tuple, whose first number is used to represent the row and the second number is used to represent the column. The most left-up grid is initially assigned with(1,1).As shown in Figure 4, the x-coordinate and y-coordinate are increased if the location of a grid shifts one position toward right and down directions respectively [31].

To guarantee that any camera sensor's sensing range can full-view cover its grid, the length of each grid is $d$ which is a predefined constant based on mobile camera full-view coverage model. In the following, we show a necessary and sufficient condition on the grid length such that the area can be full-view coverage.

**Theorem 5.1** (Grid length) Given $0 \leq h \leq r$, $45° \leq \theta \leq 90°$, in order to guarantee that the grid area satisfies full-view coverage, the length $d$ of each grid should satisfy the relationship: $d \leq \delta = 2h = r \cdot 2\sqrt{5}/5$. When $d = \delta = 2h = r \cdot 2\sqrt{5}/5$, the number of deployment cameras are minimum Where $\delta$, $h$ and $r$ are proposed in theorem 4.6.

Proof: If the grid area satisfies full-view coverage, we can obtain the relationship of grid length according to mobile camera full-view coverage proposed by the previous sections. When the length $d \leq \delta = 2h = r \cdot 2\sqrt{5}/5$, the coverage area satisfies full-view coverage. As we hope the number of selected cameras in full-view coverage is minimum, we choose the length $d = \delta = 2h = r \cdot 2\sqrt{5}/5$. As a result, the conclusion is obvious.

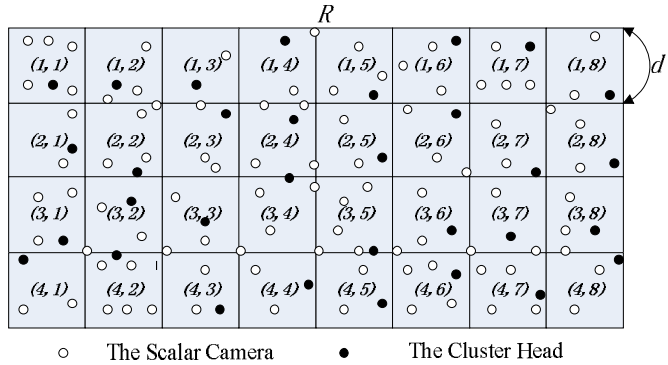

**Figure 4: The grid-based deployment strategy**

## 5.2 Grid Deployment Strategy

Since camera sensors are deployed randomly in a target field, how to move cameras with location information to form the full-view coverage is a challenging problem. We design a grid-based deployment strategy algorithm for the full-view coverage. The main idea of the algorithm 1 is as follows:

**Partition**: First, the entire region $R$ is divided into connected grids with the side length $d$, which is an $m \times n$ sized matrix. At the same time, we mark on the label for each grid. Let notation *cell(x, y)* denotes the coordinate of grid and *count(x, y)* denotes the number of camera sensors in grid *cell(x, y)*.

**Grid head**: Each camera is aware of its own location and the side length of grid, so it is sure that other cameras' location in this grid. As between camera sensors can communicate with each other within the range of *2r*, it is certainly clear that how many cameras are in this grid and cameras ID respectively. We select the optimal camera as cluster (grid) head from every grid.

**Assigning tasks**: Firstly, The grid head is responsible for collecting the information of its members, and determining cameras locations and the number of cameras in the grid. The next step job is to arrange all camera members to four vertices of the grid. (Since each camera knows its own position and the grid position, so it can calculate the coordinates of the four vertices of the grid.)

**Movement**: The grid head divides cameras with the formula: *count(x, y) / 4*, then the distribution cameras move average to four vertices of the grid. The moving orders are the left top, the right top, the left bottom, the right bottom of four vertices respectively.

**Orientation**: We elect one camera from each vertex and adjust the direction along the grid vertices. In each vertex of the grid, the selected camera can rotate direction, the other cameras keep silent. It should point to the downward for the top selected cameras of the first row, and it should point to the upward for the bottom selected cameras of the last row. For vertices on any other rows, it is necessary select one camera which points to the upward, and then we select the second camera which points to the downward. The rest of cameras keep silent. When the previous selected cameras fail, other cameras will be taken up as substitutes.

**Row Repeat**: After grids deployment in the first row is completed, other row grids begin deployment, until the end of the last row.

**Repeat column**: After the row deployment is fixed, similar operations occur in every column, until the end of the last column of the last row. The entire deployment is completed.

The algorithm 1 can be described as follows.

---

**Algorithm 1** Grid Deployment Strategy for the Full-View Coverage

---

Initialization:

    1.*R=area(m,n,d)*: The monitored area *R* is divided into grids with *m* row and *n* column and the side length *d*.

    2.*cell(i, j)*:The grid in *R* whose row is *i* and column is *j*,$1<=i<=m,1<=j<=n$.

    3.*cell -included(i, j)*:The set of cameras is included in the *cell(i, j)*.

    4.*count(i, j)*:The count of cameras is included in *cell(i, j)*.

    5.*SHij*:The grid head in the grid *cell(i, j)*.

    6.*cell-head={SH11, ……SH21, ……SHmn}*:The set of camera heads which are in the whole grids.

    7.*vertex(i, j)*:The set of vertex of *R*,$1<=i<=m+1,1<=j<=n+1$.

    8.*vertex-camera-set(i, j)*:the set of cameras which stay in *vertex(i, j)*.

Process:

    1. Select grid head from each grid *cell (i, j)* is consist of *cell-head*.

    2.for each grid head in *cell-head*

        if count of *cell -included(i, j)* >0

          distribute camera sensors to four average vertexes in *cell(i, j)*;

          the partition cameras move to the four vertex of *cell(i, j)*;

          create *vertex-camera-set* and add relevant cameras to *Vertex-camera-set(i, j)*;

        end if

      end for

    3.for each *row* in *m+1*

        3.1.if *row ==1*

            for each *j* in *n+1*

              if *count(i, j)*>0

              random select one camera in *cell-included(1,j)*;

              rotate the selected camera and set its direction along the downward

              end if

            end for

          end if

        3.2.else if *row==m+1*

            for each *j* in *n+1*

               if *count(m+1,j)*>0

              random select one camera in *cell-included (m+1,j)*;

              rotate the selected camera and set its direction along the upward

              end if

            end for

          end if

        3.3.else *1<row<m+1*

          for each *j* in *n+1*

            if *count(row, j)*>0

random select two cameras in *cell-included (row, j)*;
rotate one selected camera and set its direction along the downward
rotate another selected camera and set its direction along the upward
end if
end for
end if
end for
Return: mobile camera deployment in the area *R*.

## 6. CAMERA SELECTION FOR BARRIER COVERAGE

According to the description mentioned above, the monitored field is partitioned into many mutually connected grids. Based on the partition, we model the entire full-view covered grids and their relationship as a graph. In this graph, each node represents a full-view covered grid, and two nodes can be connected by an edge or a point if and only if they are adjacent in the original field. By doing this, we propose an algorithm to select a camera barrier. Firstly, we find all paths from one boundary to the opposite boundary on the graph, which consists of nodes that are full-view covered. The corresponding camera sensors of all nodes in the path form many camera barriers. Then we find a shortest camera barrier which contains the minimum number of camera sensors.

The details are as follows: we construct a virtual node weight graph $G= (V, E, w)$ to model all the full-view covered grids and their relationship with each other. Each full-view covered grid corresponds to a node in $V$. For any two nodes, there is an edge between them if and only if their corresponding two full-view covered grids share at least one common point or boundary. We add two virtual nodes $s$ and $t$ into this graph, which represent the left and right boundaries of field $R$ respectively. The point $s$ intersects with all grids for the left boundary of $R$. Similarly, the point $t$ intersects with all grids for the right boundary of $R$.

For each grid $v_i \in V-\{s , t\}$, $E [i]$ is equivalent to edge $v_i v_j$ , the weight $w(v_i, v_j)$ of edge $v_i v_j$ represents camera number incremental which is caused the selected grid $v_j$. In other words, we exclude the shared camera sensors between the grid $v_i$ and $v_j$. For $s$ and $t$, $w(s, v_6) =4$, $w (v_9, t) =0$. An example of this graph is shown in Figure 5. If we only cover one grid, it needs four different cameras to finish it. Considered to the adjacent or diagonal two grid nodes, the required cameras will be greatly reduced. It needs six different cameras for covering the adjacent two grids (i.e., $v_{11} v_{12}$) simultaneously. However, it needs seven different cameras for covering the diagonal two grids (i.e., $v_1 v_7$) simultaneously. In other words, when we select full-view covered grids, if the relationship is adjacent between grids, then the boundary weight between them is 2. If the relationship is diagonal between grids, the boundary weight between them is 3.

When $G= (V, E, w)$ is constructed, we can simplify the graph by removing all nodes that have only one neighbor except $s$ and $t$ since they cannot be intermediate node of any $s$–$t$ path. After all these processes, if there exist paths from $s$ to $t$, which will be actually corresponding to a series of connected grids that are all full-view covered. We find a camera barrier $C$ from all paths which requires the minimum number of mobile cameras in the region $R$. Our problem can be converted to find a $s$–$t$ path in $G$ with containing minimum number of cameras. To calculate the shortest path problem between $s$ and $t$, we execute classical Dijkstra's algorithm [35].Since many existing literatures [29, 36] on Dijkstra's algorithm can be directly applied, we will not address these issues in this paper. The Dijkstra's algorithm proposed above can be used to find a feasible solution to form a camera barrier $C$.

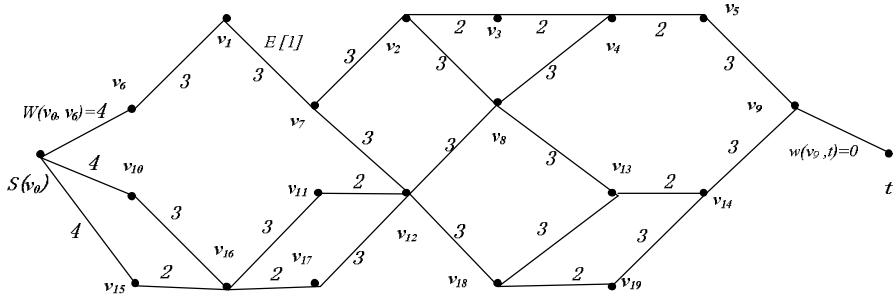

(a)The grid-based division model

(b)The relationship graph for full-view covered grids

**Figure 5: The division model and relationship graph for full-view covered grids**

As shown in figure 5, we can also realize *k*-barrier coverage except for realizing minimum camera barrier coverage. The specific method is as follows: we first need to search nodes in the graph. By calculating the number of all nodes in every column, then we select the minimum nodes *k* for all columns. The number *k* represents that the entire region can be achieved *k*-barrier coverage. It is notable that we believe that this study will open a new door to explore *k*-cameras barrier coverage.

## 7. EVALUATIONS

In this section, simulation results are presented to validate the efficiency of mobile camera barrier coverage. We first show that mobile camera full-view covered model improves cameras availability in comparison with static camera full-view coverage, and then apply this model to explore how camera parameters would affect coverage probability. Secondly, we compare the number of selected cameras under our proposed barrier coverage algorithm with that under minimum camera sensors path selection algorithm proposed in [1], and then we evaluate the number of camera sensors for constructing barrier coverage with different camera parameters.

### 7.1 Comparison with Mobile Camera Full-View Coverage

Previous works in full-view coverage mainly focused on static camera sensors, which can lead to a large number of redundant cameras which are not used properly. In this paper, we make use of mobile camera sensors to finish full-view coverage by exploiting redundant camera sensors properly. As a result, mobile camera full-view coverage is definitely more cost-effective than static camera full-view coverage, and the saving cameras' number by mobile camera full-view coverage is significant.

Firstly, we compare coverage probability of static camera full-view coverage and mobile camera full-view coverage with the following simulation settings. The monitored field *R* is a *100m×200m* rectangle region. The camera's parameters are *r=30m, a=π/3, θ=2π/3*. Cameras are deployed randomly in the target field. To avoid the boundary effect, the target field is a much larger area. We change the number of deployed cameras from *0* to *1200* to evaluate coverage probability of static camera full-view coverage and mobile camera full-view coverage. To get the probability, Extensive simulation studies have been conducted to verify if the field is full-view coverage. The coverage probability is the average value of the simulation results. As shown in figure 6(a), we can observe that coverage probability of mobile camera full-view barrier coverage is almost 1 when the number of deployed cameras is beyond *500*, while that is at least *1000* cameras for static camera full-view coverage. It demonstrates that the number of cameras required by mobile camera full-view coverage is much less than static camera full-view coverage. This result is consistent with our expectation, and the advantage of mobile camera full-view barrier coverage is even more obvious when the size of the target field is larger.

Then, we study the impact of several different max-viewing-angles *a* for mobile camera full-view coverage on coverage probability. In this experiment, the camera's parameters, the width and the length of the target area are same as previous deployment, the max-viewing-angle is *a=π/3, a=π/4, a=π/2* respectively. Figure 6(b) shows how the coverage probability varies with the number of deployed cameras under several different max-viewing-angles *a*. To verify the coverage probability, we run extensive simulation results to conduct it. Intuitively, smaller *a* requires more cameras to cover the target area. This is because the max-viewing-angles *a* is smaller, the more few target area is full-view covered by this camera. As figure 6(b) turns out, when *a = π/3*, about *500* deployed cameras can cover the target with probability approaching 1. On the other hand, more than *500* cameras are needed if *a = π/4*. Similarly, as *a = π/2*, we only need no more than *500* cameras to satisfy about 100% coverage probability.

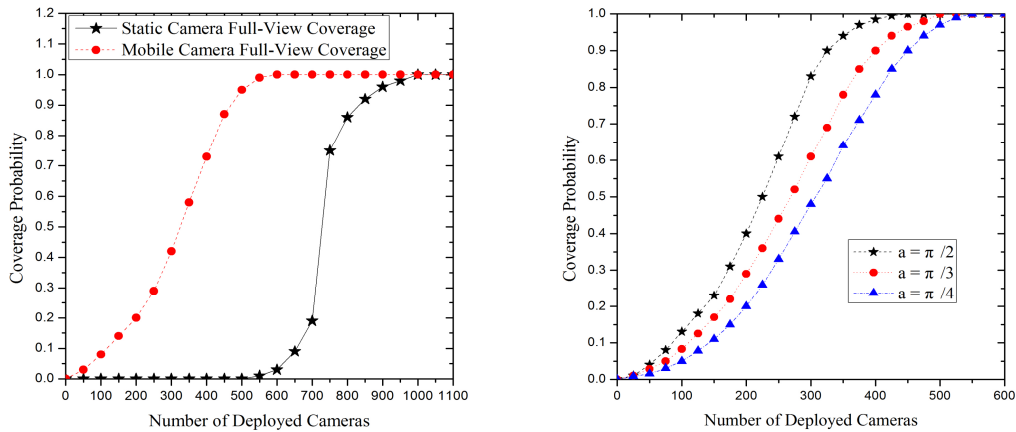

**(a) Comparison of static/Mobile camera coverage**  **(b) Comparison of max-viewing-angles *a***
**Figure 6: Coverage probability vs. Number of deployed cameras**

## 7.2 Number of Cameras for Constructing Barrier Coverage

In this section, we discuss the relationship between the number of cameras for constructing the camera barrier and the number of deployed cameras in two different scenarios. One is that we compare the number of selected cameras under our proposed barrier coverage algorithm with that under minimum camera sensors path selection algorithm. The other is that we study the impact of several different max-viewing-angles *a* on the number of cameras for constructing the barrier. Figure 7 shows how the number of cameras for constructing the camera barrier varies as the number of deployed cameras increases.

In the first scenario, the target field is *100m* in length. The width of the target field is *50m* and the number of deployed cameras will change. The camera's parameters are *r=3m, a =π/3, θ=2π/3*. We run extensive simulation to verify the experiment results. The probability is the average value of the test results. As shown in figure 7(a), the number of cameras for constructing the camera barrier selected by our algorithm is much smaller than that by the minimum camera sensors path selection algorithm. The reason is that much redundant cameras are selected to form the camera barrier in the minimum camera sensors path selection algorithm. However, the redundant cameras can be used for our algorithm reasonably. Moreover, as more cameras are deployed, there is more obvious conclusion on camera selection. Above simulation results show that our algorithm outperforms the minimum camera sensors path selection algorithm in [1].

In the second scenario, the width and the length of the target field are the same as that in the first scenario. The sensing range is *r=3m* and the field-of view angle is *θ=2π/3* and the max-viewing-angle is *a=π/3, a=π/4, a=π/2* respectively. As shown in figure 7(b), the number of cameras for constructing the barrier increases as the number of deployed cameras increases, This is because more deployed cameras implies that more cameras will be selected to cover the target area. In addition, smaller *a* implies that we need more deployed cameras to realize barrier coverage. Such behavior can be attributed to the reason that smaller *a* need more cameras to finish full-view coverage.

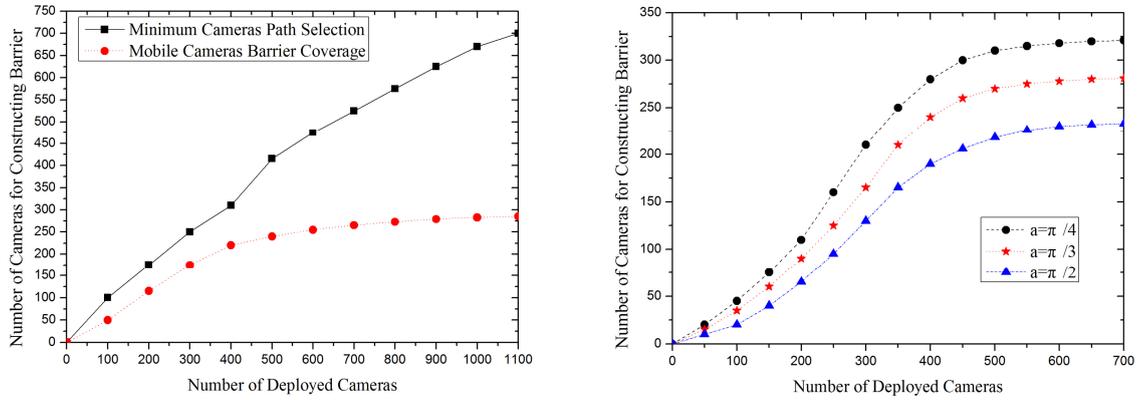

**(a) Comparison of mobile /Minimum cameras coverage**      **(b) Comparison of max-viewing-angles *a***
**Figure 7: Number of cameras for barrier construction vs. Number of deployed cameras**

## 8. CONCLUSIONS

In this paper, we have investigated barrier coverage using mobile camera sensors with grid-based deployment strategy. After the initial random deployment, we first propose a novel full-view covered model based on mobile camera sensors. With this model, we divide the original continuous target space into connected discrete grid-based spaces which can satisfy the condition of full-view coverage, and deploy mobile camera sensors for each grid. Then we introduce the concept of virtual node to construct a weighted directed graph, which is used for exploiting geographical relationships among the full-view covered grids. Finally, we employ Dijkstra's algorithm to obtain a shortest camera barrier from the source node to the destination node on the weighted directed graph. Our simulation results show that our algorithm only need the minimum number of mobile camera sensors while maintaining the same coverage performance in comparison with previous algorithms. Therefore, our solution can be used to better improve the performance of barrier coverage by mobile camera sensors, and the results also provide a guideline for barrier coverage of large scale mobile camera sensor networks.

## 9. ACKNOWLEDGEMENTS

This work was partly supported by the Natural Science Research Projects of Chuzhou University under grant 2012kj005B.

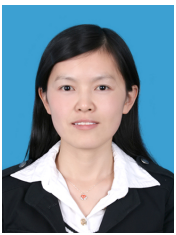

**Xiaolan Liu (刘晓兰)** received her Master degree in information and Communication Engineering from Dalian Maritime University in 2009. Since 2009, he has been with School of Computer and Information Engineering at Chuzhou University. Her research interests wireless camera sensor networks and barrier coverage.

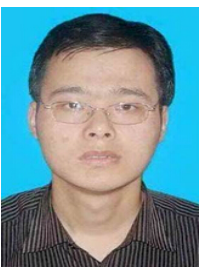

**Bin Yang (杨斌)** received his B.S. and M.S. Degrees both from Shihezi University, China, in 2004 and from China University of Petroleum, Beijing Campus, in 2007, respectively. He is currently a Ph.D. candidate at the School of Systems Information Science, Future University Hakodate, Japan and is also faculty member at the School of Computer and Information Engineering, Chuzhou University, China. His research interests include performance modeling and evaluation, stochastic optimization and control in wireless networks, LTE-A and 5G networks.

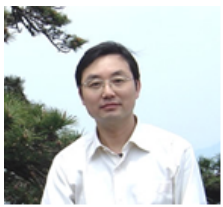

**Guilin Chen (陈桂林)** received his B.S. and M.S. Degrees both in Mathematics from Anhui Normal University, china, in 1985 and in Computer Application from Hefei University of Technology, China in 2002. He is currently a Professor in the School of Computer and Information Engineering at Chuzhou University, China. His main research interests include cloud computing, Internet of Things and big data.